# Forming Three-Dimensional Colloidal Structures Using Holographic Optical Tweezers

C.R. Knutson and J. Plewa

*Arryx, Inc. 316 N. Michigan Ave Chicago IL 60601*

*(312) 726-6675 ext. 248*

*(312) 726-6675 ext. 233*

*(312)726-6652 (fax)*

cknutson@arryx.com

jplewa@arryx.com

A method for forming permanent three dimensional structures from colloidal particles using holographic optical trapping is described. Holographic optical tweezers (HOT) are used to selectively position charge stabilized colloidal particles within a flow cell. Once the particles are in the desired location an electrolyte solution is pumped into the cell which reduces the Debye length and induces aggregation caused by the van der Waals attraction. This technique allows for the formation of three dimensional structures both on and away from the substrate that can be removed from solution without the aid of critical point drying. This technique is inexpensive, fast, and versatile as it relies on forces acting on almost all colloidal suspensions.

**I. INTRODUCTION**

Structures formed from colloidal particles hold great promise for applications that reach across a wide variety of fields. These structures typically consist of particles with diameters of a few nanometers to a few microns and can be formed from a wide variety of materials with specific chemical morphology. This tunability gives researchers the capability to form devices that can exhibit interesting optical,[1,2,3] electronic,[4,5] and potentially magnetic behavior.

Typical formation of two- and three-dimensional colloidal structures relies on self-assembly driven techniques such as template-assisted self-assembly or various field-driven techniques.[6,7,8,9,10,11] Such techniques generally produce large scale structures in a short amount of time but lack the capability to form long-range defect-free structures. In addition, forming complex yet regular colloidal crystals composed of two or more colloid types with different diameters and/or compositions is quite difficult.

Recently, researchers have demonstrated that structures formed from colloidal particles can be made by positioning particles in a polymer gel solution with multiple optical traps.[12, 13] Once the gel sets the particles remain fixed in place. This technique can produce three-dimensional structures but requires that the structure be formed in a gel matrix. Additionally, other researchers have demonstrated that relatively large two-dimensional structures could be formed by positioning charged colloidal particles on an oppositely charged substrate with optical traps generated by an acousto-optical deflector system.[14] These structures could also be removed from solution with the aid of critical point drying.[14] This technique has great potential for the formation of devices that rely on precisely positioned colloidal particles. However, the technique is generally limited to the formation of two-dimensional colloidal structures on a substrate.

In this letter we report a technique that allows for two- and three-dimensional structures to be formed on a substrate or in solution from charged stabilized colloidal particles using HOT. The use of HOT allows relatively large numbers of particles composed of a variety of substances to be precisely positioned without the introduction of foreign objects into a sample cell.[15] The assembly technique relies on altering interaction potentials that exist in almost all colloidal suspensions, namely the Coulomb and van der Waals interaction which makes the technique widely applicable to many systems. Finally, we demonstrate that structures built with this can withstand the forces associated with removing the structure from solution without the aid of critical point drying. This technique opens a new route to form large two- and three-dimensional colloidal structures composed of a wide variety of materials that are capable of being removed from solution for further use.

## II. METHODS AND RESULTS

In our configuration the holographic optical traps are generated using the Arryx Bioryx 200™ ® system utilizing a 532 nm continuous wave laser (Spectra Physics Millennia V) on an inverted microscope (Nikon TE-200). The system allows the formation of up to 200 hundred traps with independent control of the x, y, and z position. In these studies a 60X high numerical aperture (n.a.=1.4) oil immersion objective is used. The flow cell is created by affixing the input and output flow tubes (Tygon tubing, OD=0.40", ID=0.07") to a 50 mm by 22 mm #1 coverslip with Devcon Five Minute Epoxy Gel and then placing a standard glass slide on top of the epoxy well. The input tube is connected to a 3 mL syringe and a syringe pump (WPI SP2001) is used to control the flow rate of the electrolyte solution into the sample cell. Figure 1 is an illustration of the assembly sample cell.

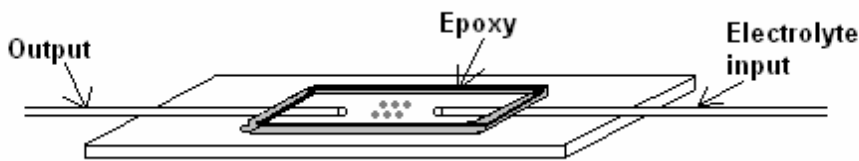

FIG. 1. Illustration of the experimental sample cell used for assembly. The electrolyte input tube is connected to a syringe pump to control the flow rate of electrolyte into the sample cell. Precise control of the flow rate ensures trapped particles are not forced free from the optical traps by viscous drag forces.

The assembly process begins by filling the sample chamber with a dispersion of commercially available monodisperse silica colloid (Bangs Laboratories #SS04N/5569) with diameter of 2.34 um and sealing the sample cell with epoxy. The colloid is used as is without any surface modification and dispersed in filtered water (Barnstead Nanopure – pH adjusted to 7.0 with 0.05 M NaOH) to produce a concentration of approximately $1 \times 10^6$ particles/mL. The sample cell is placed on the microscope stage and the particles are allowed to sediment to the coverslip surface. This concentration produces a two-dimensional packing fraction of approximately 2% that achieves an excellent balance between availability of free particles and open space to form structures.

Particles are then trapped with optical tweezers and held in place away from each other and the walls of the flow cell. Typically we acquire 10 to 50 particles in one assembly step with an average single trap power of 4 to 20 mW. These trap powers ensure that trapped particles can withstand viscous drag forces associated with introducing the electrolyte solution into the cell. Once the particles have been trapped a 0.2 M NaCl (Sigma Aldrich #S-7655) solution is flowed into the sample cell. Although we find that diffusion alone can produce the desired effect by simply introducing salt into the input syringe, flowing electrolyte solution into the sample cell ensures that the entire cell volume consists of the same concentration of electrolyte as the source. We use a flow rate of approximately 0.1 mL/min depending on the number of particles trapped and flow one to two times the sample cell volume worth of electrolyte into the cell. Un-trapped particles in the field of view act as indicators of the aggregation process. Once the thermal motion of these indicator particles stops, trapped particles are brought into contact with the coverslip or each other by either placing individual particles on the surface or in contact with neighboring particles. Additionally, entire groups of particles can be brought into contact simultaneously with the substrate by adjusting the trapping plane of groups of particles. Most particles aggregate within a few seconds of being brought into contact with the substrate or each other. However, several particles still display Brownian motion for a few seconds before adhering to the substrate or each other. This may be caused by imprecision in the location of the traps with respect to the adhesion point or perhaps variability in local electrostatic conditions.

The aggregation behavior can be understood with the celebrated DLVO theory of colloidal stability.[16] Under the prepared conditions the silica particles and the glass surfaces develop a negative surface charge principally caused by the disassociation of terminal silanol groups. At low electrolyte concentrations and pH higher than the isoelectric point of silica this surface charge prevents particle flocculation and aggregation to the glass surfaces. Under these conditions, the Debye length is typically of order 1 µm at which length the van der Waals interaction is much less than the repulsive zero frequency electrostatic interaction.[16] However, once an electrolyte is introduced into the sample cell, the Debye length decreases as surface charge is screened by ions in solution which allows particles to approach like-charged surfaces within small separations where the strength of the van der Waals interaction increases. This then allows the particles to irreversibly aggregate.[16,17] Our calculations show that the potential energy barrier associated with repulsive like-charge interaction should be sufficiently suppressed at these electrolyte concentrations.

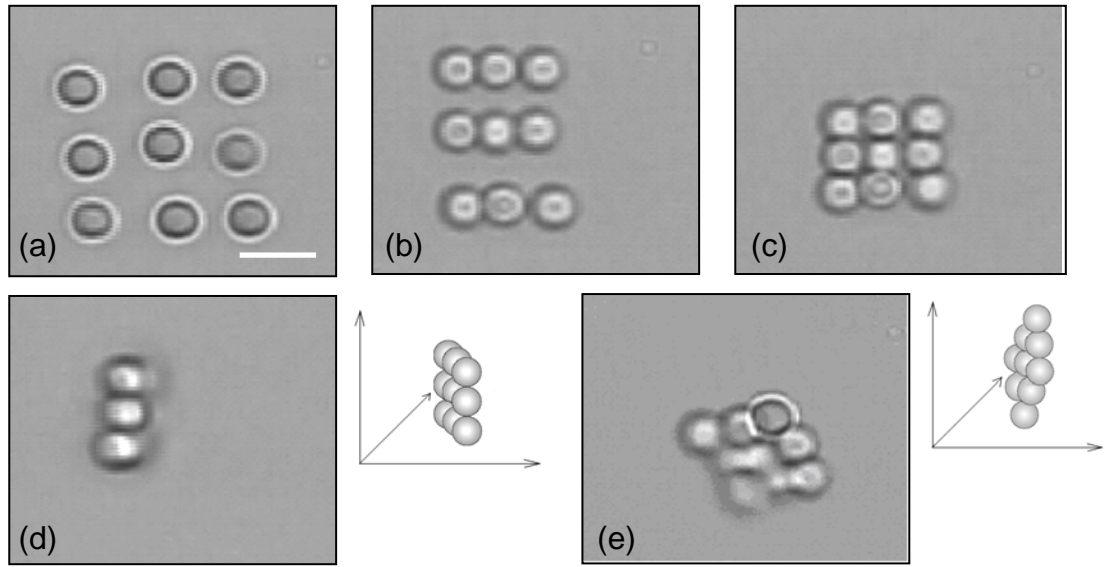

FIG. 2. Assembly of a 3×3 colloidal crystal in free space using HOT. a) Nine particles are trapped and separated from each other and the glass coverslip. A 0.2 M NaCl solution is then introduced into the sample cell. b) Particles are then joined to form three 3×1 arrays. c) The complete structure is formed by combining the separate arrays. d) The array is rotated 90° above the substrate using four optical traps. e) The structure is rotated once more by changing the trapping plane of three optical traps and deposited on the substrate to produce a 3×3 crystal in a diamond-like orientation. No optical traps are present in the final image. The scale bar is 5 μm.

Figure 2 is a collection of images illustrating the assembly process of a 3×3 colloidal crystal formed in solution away from the substrate surface, rotated twice, and then deposited on the substrate. Particles are first trapped and held apart from the substrate and each other. Particles are separated from neighboring surfaces by approximately 2 μm. A 0.2 M electrolyte solution is then introduced and the particles are combined in a 2D simple square lattice pattern until a 3×3 crystal is suspended above the substrate by approximately 10 μm. The crystal is rotated 90° about an axis parallel to the coverslip using four optical traps and then rotated again by 45° before being deposited on the substrate. In this case only one particle is in contact with the coverslip surface. The contact is sufficiently rigid to prevent visible diffusive rotation of the crystallite. Tweezing with single or multiple traps with a net power of approximately 0.2 W has no clear effect on the structure once it is adhered to the glass substrate.

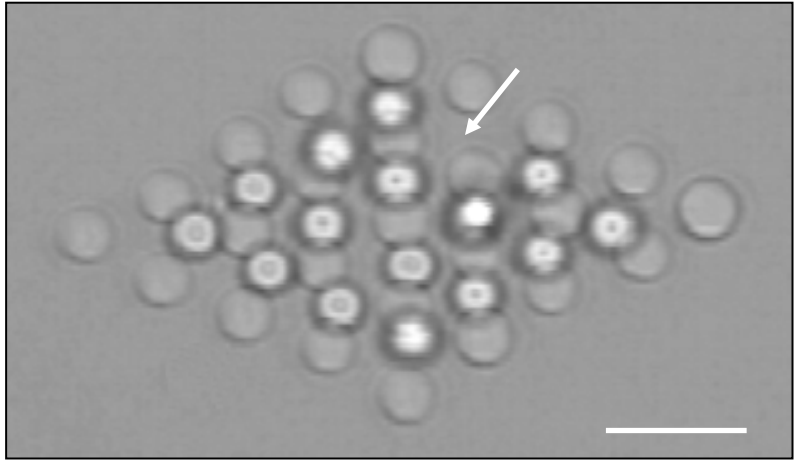

FIG. 3. A three-dimensional crystal composed of 2.34 $\mu$m silica particles. Particles are first trapped and each layer of the crystal organized into a hexagonal pattern at two different focal lengths. A 0.2 M NaCl solution is then introduced and the first layer of particles deposited on the substrate by adjusting the focal length of these traps. The second layer is then deposited particle by particle. No optical traps are present in this image. The arrow indicates a vacancy in the second layer. The scale bar is 5 $\mu$m.

Figure 3 is a brightfield image of a larger (n=40) two layer colloidal crystal formed on the glass coverslip substrate. In this case two layers of trapped particles are arranged into a hexagonal pattern at different trapping planes and then the electrolyte solution is introduced. The initial layer of particles is deposited by reducing the trapping plane of the collection of traps until all of the particles in the first layer are in contact with the coverslip. After this initial layer is deposited individual particles are positioned within the lattice to form a second layer. After each particle is deposited the optical traps are removed from the structure. No optical traps are present in the final image.

Again, once the particles are adhered to the substrate or each other, tweezing has no apparent effect on the position of the particle even at trap powers approaching 0.2 W. This is to be expected since the energy of the applied optical trap (~ 50 $k_bT$) is much less than typical van der Waals energy (~ hundreds to thousands of $k_bT$) at nanometer length scales.

Additionally, we demonstrate that this technique can be used to build a three-dimensional colloidal structure composed of two differently sized colloids. In this case 2.34 $\mu$m and 4.50 $\mu$m (Bangs Laboratories #SS059/4908) silica spheres are dispersed together in the sample cell, trapped, and then organized as in previous experiments. A 0.2 M NaCl solution is flowed into the sample cell and a single layer of 4.50 $\mu$m silica is then assembled into a two-dimensional hexagonal close packed lattice on the glass coverslip. Five 2.34 $\mu$m silica particles are then deposited over the first layer of colloid. Figure 4a is a brightfield image of the complete structure formed from two populations of differently sized silica spheres. The capability to build structures of two differently sized colloids should be useful for the formation of complex photonic crystals. In addition, this technique should give rise to the capability of building structures composed of multiple colloid types where one colloid type may be dissolved by a series of washings. This may open the possibility to form precisely positioned defects in photonic crystals for use as waveguides, for example. In order to form true crystals that display a fully developed three-dimensional bandgap a crystal composed of several layers is needed.  Although we only form structures consisting of two layers in this report we are currently developing techniques that should allow for the formation of crystals consisting of at least 50 layers of colloidal spheres with diameters as small as 400 nm.

      Finally, many of these structures can be removed from solution without the aid of critical point drying.  To demonstrate this we seal one side of the chamber with hot glue rather than epoxy.  Once the structure is formed the hot glue is liquefied using a soldering iron and one side of the chamber is exposed to atmosphere which allows the aqueous phase to evaporate from the sample cell.  This process takes several hours depending on the proximity of the structure to the open edge of the sample cell and the thickness of the sample chamber.  Figure 4b is a brightfield image of the structure in Figure 4a after the aqueous phase has been removed.  We observe that three-dimensional structures bound to another surface by a single connection or crystals with large lattice constants typically undergo minor deformation upon drying.  This observation is understandable given the low number of particle/particle and particle/substrate adhesion points and the large surface area of structure.

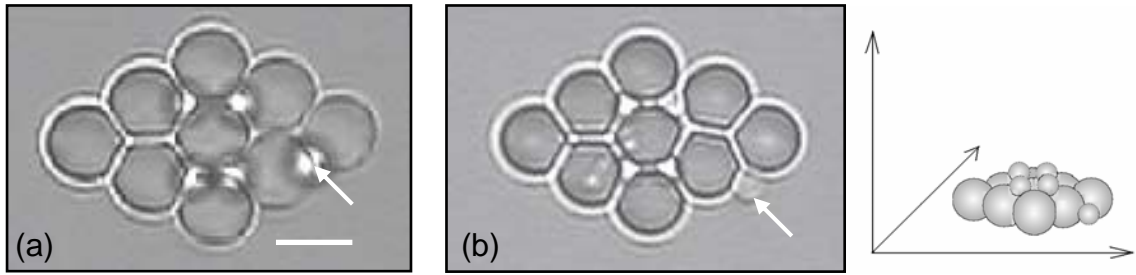

FIG. 4. A crystal formed from 4.50 and 2.34 µm silica and then removed from solution. a) An initial layer of 4.50 µm silica is deposited in a HCP pattern and four 2.34 µm silica particles are deposited at the junction of three of the larger particles. In addition, one smaller particle is placed at the junction of two particles. b) The same structure after the aqueous phase was allowed to evaporate. The arrow indicates a 2.34 µm particle that shifted position during the removal of the aqueous phase. The scale bar is 5 µm.

## III. CONCLUSIONS

We have developed a technique that allows the construction of two- and three-dimensional structures composed of multiple colloid types to be formed on or away from a substrate. The technique relies upon forces acting on many colloidal dispersions making it applicable to a wide variety of colloid types and compositions. In addition, many of these structures can be removed from solution without the aid of critical point drying. This technique may find uses in the formation of photonic crystals, colloidal electronics, and bioengineered materials.

## ACKNOWLEDGEMENTS

The authors would like to thank Anthony Dinsmore and David Grier for their critical reading of this manuscript.

FIG. 1. Illustration of the experimental sample cell used for assembly. The electrolyte input tube is connected to a syringe pump to control the flow rate of electrolyte into the sample cell. This ensures trapped particles are not forced free from the optical traps by viscous drag forces.

FIG. 2. Assembly of a 3×3 colloidal crystal in free space using HOT. a) Nine particles are trapped and separated from each other and the glass coverslip. A 0.2 M NaCl solution is then introduced into the sample cell. b) Particles are then joined to form three 3×1 arrays. c) The complete structure is formed by combining the separate arrays. d) The array is rotated 90° above the substrate using four optical traps. e) The structure is rotated once more by changing the trapping plane of three optical traps and deposited on the substrate to produce a 3×3 crystal in a diamond-like orientation. No optical traps are present in the final image. The scale bar is 5 μm.

FIG. 3. A three-dimensional crystal composed of 2.34 μm silica particles. Particles are first trapped and each layer of the crystal organized into a hexagonal pattern at two different focal lengths. A 0.2 M NaCl solution is then introduced and the first layer of particles deposited on the substrate by adjusting the focal length of these traps. The second layer is then deposited particle by particle. No optical traps are present in this image. The arrow indicates a vacancy in the second layer. The scale bar is 5 μm.

FIG. 4. A crystal formed from 4.50 and 2.34 μm silica and then removed from solution. a) An initial layer of 4.50 μm silica is deposited in a HCP pattern and 2.34 μm silica is deposited at the junction of three of the larger particles. In addition, one smaller particle is placed at the junction of two particles. b) The same structure after the aqueous phase was allowed to evaporate. The arrow indicates a 2.34 μm particle that shifted position during the removal of the aqueous phase. The scale bar is 5 μm.